\documentclass[atmosphere,article,accept,pdftex,moreauthors]{Definitions/mdpi} 
\usepackage{url}
\usepackage{rotating}
\usepackage{array}       
\usepackage{tabularx}
\usepackage{xcolor}
\firstpage{1} 
\pubvolume{16}
\issuenum{1}
\articlenumber{937}
\pubyear{2025}
\copyrightyear{2025}
\externaleditor{ Sergey Pulinets and Alexei Dmitriev } % More than 1 editor, please add `` and '' before the last editor name
\datereceived{30 June 2025} 
\daterevised{30 July 2025} % Comment out if no revised date
\dateaccepted{4 August 2025} 
\datepublished{5 August 2025} 
\hreflink{https://doi.org/10.3390/atmos16080937} % If needed use \linebreak
\Title{Investigating a Characteristic Time Lag in the Ionospheric F-Region’s Response to Solar Flares}

\TitleCitation{Investigating a Characteristic Time Lag in the Ionospheric F-Region’s Response to Solar Flares}

\Author{Aisling N. O'Hare 
 *\orcidA{}, Susanna Bekker, Harry J. Greatorex, and Ryan O. Milligan }

\AuthorNames{Aisling N. O'Hare, Susanna Bekker, Harry J. Greatorex and Ryan O. Milligan}

\isAPAStyle{%
       \AuthorCitation{O'Hare, A. N., Bekker, S. Z., Greatorex, H. J., \& Milligan, R. O.}
         }{%
        \isChicagoStyle{%
        \AuthorCitation{O'Hare, Aisling N., Bekker, Susanna., Greatorex, Harry J., \& Milligan, Ryan O.}
        }{
        \AuthorCitation{O'Hare, A.N.; Bekker, S.; Greatorex, H.J.; Milligan, R.O.}
        }
}

\address[1]{%
 Astrophysics Research Center, School of Mathematics and Physics, Queen's University Belfast, University Road, Belfast BT7 1NN, UK
; s.bekker@qub.ac.uk (S.B.); h.greatorex@qub.ac.uk (H.J.G.); r.milligan@qub.ac.uk (R.O.M.) 
\\
}
\corres{\hangafter=1 \hangindent=1.05em \hspace{-0.82em} Correspondence: aohare32@qub.ac.uk}

\abstract{X-ray and EUV solar flare emission cause increases in the Earth's dayside ionospheric electron density. While the response of the lower ionosphere to X-rays is well studied, the delay between EUV flare emission and the response of the ionospheric F-region has not been investigated. Here, we calculate the delays between incident He~{\sc{II}} 304\,\AA{} emission, and the TEC response for 10 powerful solar flares, all of which exhibit delays under 1\,minute. We assess these delays in relation to multiple solar and geophysical factors, and find a strong negative correlation {($\sim$$-$0.85)} between delay and He~{\sc{II}} flux change and a moderate negative correlation {($\sim$$-$0.55)} with rate of increase in He~{\sc{II}} flux. Additionally, flare magnitude and the X-ray-to-He~{\sc{II}} flux ratio at peak He~{\sc{II}} emission show strong negative correlations with delay {($\sim$$-$0.80 and $\sim$$-$0.75, respectively)}. We also identify longer delays for flares occurring closer to the summer solstice. These results may have applications in upper-ionospheric recombination rate calculations, atmospheric modelling, and other solar--terrestrial studies. We highlight the importance of incident EUV and X-ray flux parameters on the response time of the ionospheric electron content, and these findings may also have implications for mitigating disruptions in communication and \mbox{navigation systems.} }


\begin{document}
\section{Introduction}
\label{intro}

Solar flares release significant amounts of energy in the form of emission spanning the entire electromagnetic spectrum. Flare emission directed earthward can cause an increase in the electron density of different layers of the Earth's ionosphere \cite{mitra1974ionospheric, deAbreu2019}. The entire dayside ionosphere is primarily affected by X-ray and extreme ultraviolet (EUV) flare emission \cite{tsurutani2009brief}  that cause an increase in ionisation and molecular dissociation of the atmospheric components at different altitudes \cite{wan2005gps, qian2011variability}. It is generally accepted that EUV (100–1000\,\AA{}) flare emission affects the middle (E-region; 90–120\,km) and upper (F-region; >120\,km) {ionospheric altitudes} where the electron density is high. Softer X-rays (10–100\,\AA{}) are primarily absorbed in the E-layer, by O$_2$ and N$_2$. In the F-layer, EUV photons are the dominant source of the ionisation of atomic oxygen (O). Higher energy X-ray photons (<10\,\AA{}) can penetrate to the lower lying portion of the ionosphere (D-region; 60–90\,km) and ionise the main neutral components at these altitudes, O$_2$ and N$_2$ \cite{mitra1974ionospheric}. Subsequently, the sudden increase in solar irradiance during solar flares induces compositional changes across the entire dayside ionosphere of the Earth, which can potentially influence communication and navigation systems.

The remote sensing method for calculating TEC derived from the Global Positioning System (GPS) has become a popular and significant tool for analysing fluctuations in ionospheric electron density \cite{Zhang2011, MonteMoreno2014, Curto2019}. GPS consists of a constellation of {31} satellites arranged in six orbital planes around the Earth, at an altitude of approximately \mbox{20,200 km \cite{Davies1997}}. These satellites transmit signals at two primary frequencies, f1 = 1575.42 MHz and \mbox{f2 = 1227.60 MHz}, which are used to study ionospheric effects. The majority of the electron density in the ionosphere is located within the F-region, so ionisation of this layer by EUV is believed to be the dominant driver of fluctuations in TEC measurements (e.g., \cite{Zhang2011}). Empirical and physics-based modelling studies have identified EUV wavelengths below 350\,\AA{} as the most effective in producing F-region ionisation, with He\,\textsc{ii} 304\,\AA{} highlighted as a particularly influential line (e.g., \cite{Nishimoto2023,Le2007,watanabe2021model}). A study by \citet{Mahajan2010} reported a strong correlation between peak $\Delta$TEC and $\Delta$EUV (26–34\,nm) flux across 10 flare events, a finding supported by similar results in other flare case studies (e.g., \cite{Zhang2011}).
Importantly, rapid increases in TEC can potentially impact the accuracy of global navigation satellite system (GNSS) positioning {\cite{tiwari2009effect}}. GNSS positioning is critical in navigation and safety, making it vulnerable to disruptions caused by variations in TEC \cite{icao1996international}.

Changes in electron density in the ionospheric D-region can be probed through observing very low frequency (VLF) radio-wave signals (3–30 kHz) that propagate in the waveguide between the Earth’s surface and the D-region with relatively low attenuation of about 2–3 dB/Mm \cite{Davies1990}. These signals are perturbed by the changes in electron density in the D-region \cite{thomson2001solar, raulin2013response, hayes2021solar, nina2021modelling, bekker2023influence} and in extreme cases can cause radio blackouts. VLF propagation studies have been utilised extensively in the past to assess the impacts of solar flares on the Earth’s lower ionosphere \cite{Pant1993, thomson2001solar, mcrae2004solar, Thomson2005, Selvakumaran2015}. There are dozens of VLF receivers and transmitters across the globe that continuously measure the variations of VLF signals. Most notably, these measurements have been combined with theoretical models to estimate the electron-density profile in the D-region in response to solar flares \cite{wait1964characteristics, Bekker2020}. 

In recent years, VLF observations of solar flares have been used to assess the time delay between the peak of incident X-ray flux and the peak of the VLF amplitude response. This time delay is known as ionospheric ‘sluggishness' \cite{appleton1953note, valnicek1972x} or ‘relaxation time' \cite{mitra1974ionospheric}, and is an {inherent} property of the ionosphere that is determined by the balance between ionisation and recombination processes in the medium, as well as environmental parameters such as the solar zenith angle, the ionospheric altitude, latitude, background solar and magnetic activity. \citet{basak2013effective} reported a range of time delays between 1–8\,\AA{} X-ray peak and VLF response for 22 flares from $\sim$1–6\,minutes. Similarly, \citet{hayes2021solar} calculated time delays between both 0.5–4\,\AA{} and 1–8\,\AA{} X-rays and VLF amplitude peaks for over 300 flares of various classes, with average delays of 3.4\,minutes and 1.7\,minutes respectively. On the other hand, \citet{chakraborty2020numerical} reported delay values as low as 16\,s.  When combined with estimates of the electron density profile, the measured time delay can be used to estimate the effective recombination coefficient of the lower ionosphere \cite{igman2007, basak2013effective, Hayes2017}.

For the time delay (D) between incident EUV flare emission and the F-region response, only two single reports have been made of such a delay by \citet{bekker2024influence} and \citet{ohare2025quasi}; measuring 60 s and 30 s, respectively. \citet{bekker2024influence} identified a $\sim$1 min delay between flare-induced EUV peaks and the broader $\Delta$TEC response. \citet{ohare2025quasi} presented the first observational evidence of a $\sim$30 s delay between synchronised, small-scale pulsations in individual solar EUV emission lines and corresponding fluctuations in ionospheric TEC. One of the potential reasons for the lack of such reports are the challenges associated with this delay measurement. To measure the delay between EUV emission and TEC response, we must select moments of elevated EUV emission before X-rays have risen significantly. This challenge does not exist for the D-region as only X-ray emission will cause fluctuations in VLF measurements during solar flares. Additionally, even with a lower X-ray profile, gradual increases in EUV flare emission will mean the TEC enhancement is too gradual to confidently select the moment of response that aligns with the peak in incident emission. Therefore, measurement of this delay is intricate and requires multiple criterion to ensure robust detection.

In this paper, we present a study of 10 solar flares of magnitude >M5 observed by the Solar Dynamics Observatory (SDO; \cite{pesnell2012solar}) Extreme Ultraviolet Variability Experiment (\mbox{EVE; \cite{woods2012extreme}}) instrument from 2011--2013. The time delays between observed peaks in the He~{\sc{II}} 304\,\AA{} flare emission line and the response in TEC were calculated and assessed in relation to the overall change in He~{\sc{II}} 304\,\AA{} irradiance, the rise rate of the incident He~{\sc{II}} 304\,\AA{} irradiance, the solar zenith angle of the GNSS stations used, the season, the flare class, and the ratio of X-ray to He~{\sc{II}} 304\,\AA{} irradiance. This is the first multi-flare study of such a delay, in addition to being the first study to assess the relationship between this delay and various geophysical and solar factors. In Section \ref{sec2} of this article we outline the assessment and selection of the solar and ionospheric data used in this study. Section \ref{sec3} outlines the methods used to process the TEC data, calculate delays, and related parameters. The results of this study are presented in Section \ref{sec4}, and Section \ref{sec5} discusses their implications \mbox{and origins. }

%%%%%%%%%%%%%%%%%%%%%%%%%%%%%%%%%%%%%%%%%%
\section{Observations and Data Selection}\label{sec2}

\subsection{Solar Observations}\label{solobs}
To calculate the delay between solar emission and TEC response, the emission line responsible for the majority of the ionospheric electron increase was analysed. According to theoretical models of the Earth's ionosphere \cite{Solomon2005, watanabe2021model}, He~{\sc{II}} 304\,\AA{} is the most geoeffective line of the solar spectrum and is a dominant driver of ionospheric ionisation. He~{\sc{II}} 304\,\AA{} is predominantly emitted from the chromospheric transition region, and it contributes more energy than any other single emission line in the EUV range during a flare \cite{Woods2011}. The He~{\sc{II}} 304\,\AA{} emission line was observed at 10\,s cadence by the SDO/EVE instrument's MEGS-A channel that operated between 2010–2014. Therefore, to select the flares used for this study, we first generated a flare list of events during this time period of magnitude >M5 using Sunpy's \cite{sunpy_community2020} Fido function to query the Heliophysics Event Knowledgebase (HEK) \cite{hurlburt2012heliophysics} catalogue of GOES (Geostationary Operational Environment Satellite) flare events. This provided us with the peak X-ray flux, event start, peak, and end times, and the location of the flare on the solar disk was retrieved from the SolarSoft 
 Latest Events Archive (accessed on 1 April 2025 from \url{http://www.lmsal.com/solarsoft/latest_events_archive.html}).   
The He~{\sc{II}} 304\,\AA{} emission during the flares returned by the HEK query were manually inspected for ‘sharpness'---a quality identified by a clear defined peak reached in a short period of time relative to the overall flare time. This criterion was essential to ensure the associated TEC response was not too gradual to reliably identify the peak. Furthermore, flares deemed ‘sharp' enough were then assessed based on the timing of the X-ray increase relative to the He~{\sc{II}} 304\,\AA{} peak. X-ray flare emission typically peaks after the peak in He~{\sc{II}} 304\,\AA{}, so in order to distinguish the effect of He~{\sc{II}} 304\,\AA{} on TEC, those flares where the X-rays peaked within 1\,minute after the peak in He~{\sc{II}} were discarded. This left a flare list of 57 flares.

\subsection{Ionospheric Data}\label{ionodata}
\begin{figure}[H]

    \includegraphics[width=0.7\linewidth]{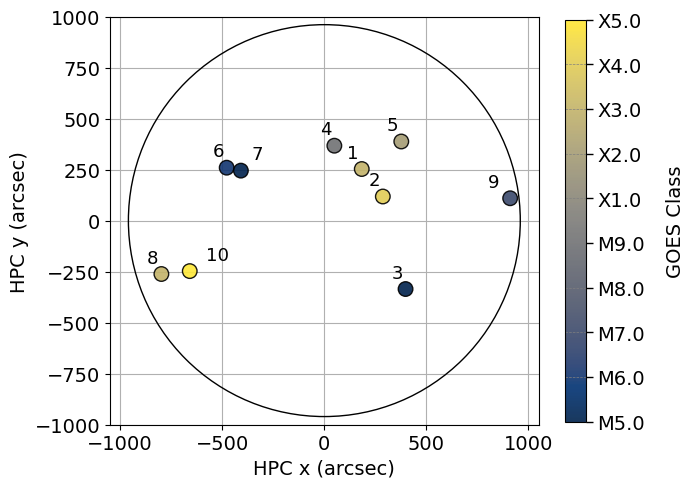}
    \caption{Map of the solar disk with axes showing Helioprojective Cartesian coordinates in arcseconds. The points on the map denote locations of the flares used in the study, with the colour indicating their GOES class.}
    \label{disk}
\end{figure}

\begin{figure}[H]

% \begin{adjustwidth}{-\extralength}{0cm}

\includegraphics[width=0.95\linewidth]{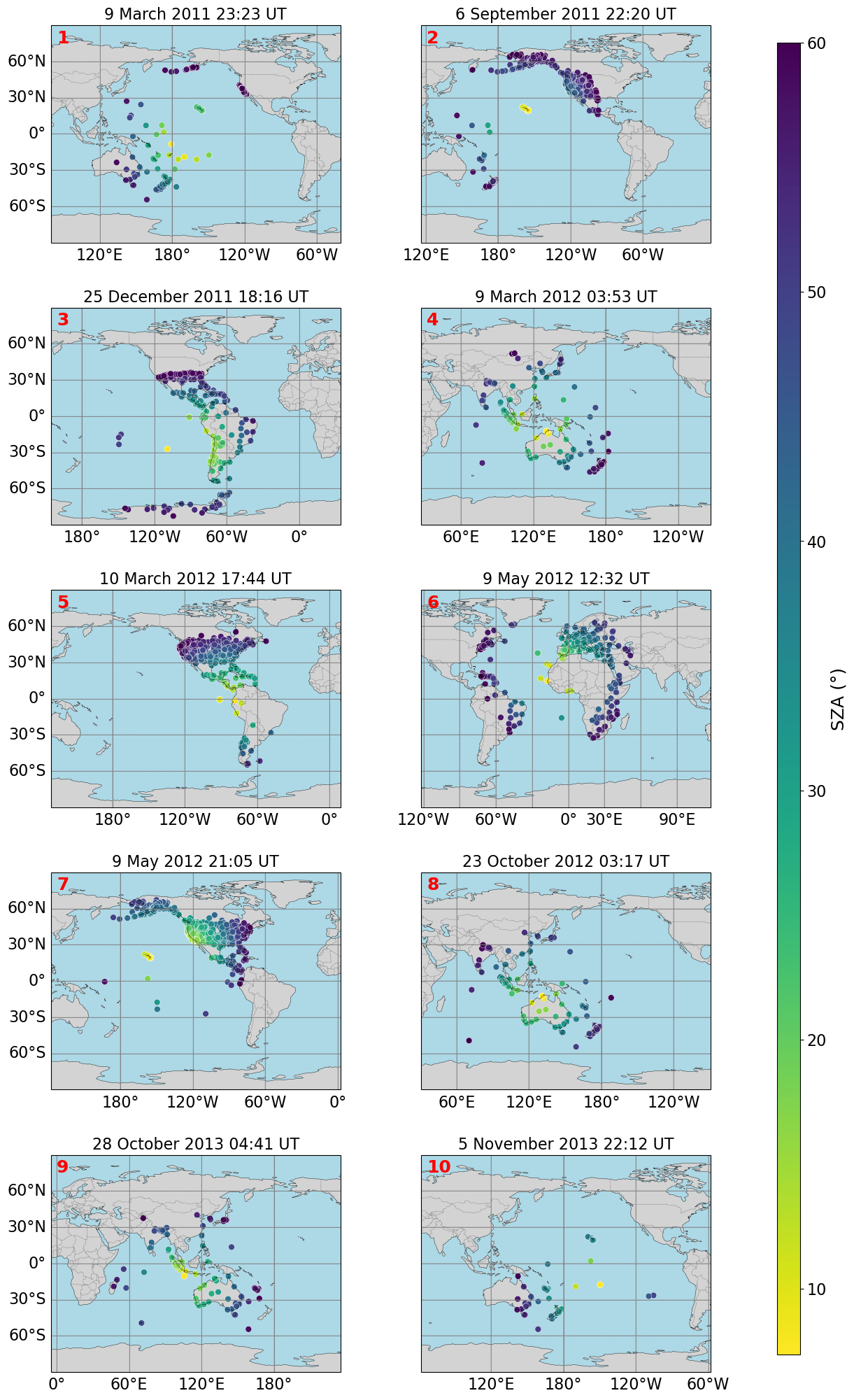}
\caption{Maps showing the locations of GPS stations in the SOPAC network for each flare in this study. The red numbers 1--10 correspond to the flare numbers listed in Table \ref{Tablesolar}. The colour denotes the solar zenith angle of each station at the peak time of each flare.}
\label{maps}
\end{figure}

For the initial 57 flares with ‘sharp' He~{\sc{II}} lightcurves, TEC variations were derived using GPS data (15\,s cadence) from the Scripps Orbit and Permanent Array Center (SOPAC) network stations. For further information on how GPS satellites operate, see \citet{HofmannWellenhof1992}. For each flare, we selected only ground-based stations where the solar zenith angle at the peak flare time was less than 60$^{\circ}$. This criterion ensures that stations were sufficiently illuminated, which is important as enhanced solar emission causes increased electron density in the illuminated part of the Earth's ionosphere. On average, 4–5 GPS satellites are simultaneously observed by each ground-based station. {We did not exclude satellites based on elevation angle, as tests showed it does not affect the peak response time of the medium studied in this work.} The slant TEC curves were then detrended and averaged (see Section \ref{detrsection} for methodology). These average detrended TEC profiles were then cross-examined with concurrent He~{\sc{II}} 304\,\AA{} and X-ray 1–8\,\AA{} lightcurves. Flares were retained for analysis provided there was a distinct TEC response to He~{\sc{II}} emission, permitting confident delay measurement, and that there was a clear separation between the TEC responses to He~{\sc{II}} and X-ray emissions. This selection process yielded 10 flares with measurable and distinct ionospheric responses to He~{\sc{II}}. Table \ref{Tablesolar} summarises some of the properties of these events, including dates, peak times, GOES classifications, and Helioprojective Cartesian (HPC) coordinates. Figure \ref{disk} displays the flares' positions on the solar disk, with the colour bar indicating the GOES class. The locations on Earth of the selected GPS stations for each of the final 10 flares can be seen in Figure \ref{maps}, where colour represents the solar zenith angle (SZA) of each station at the peak of the flare. {The average SZA, latitude, and longitude of the GPS stations, and number of satellite-station pairs used when calculating TEC are listed in Table \ref{Tablegps}.}

\begin{table}[H]
\caption{Summary of the number, date, X-ray peak time, GOES class, and Helioprojective Cartesian coordinates of the flares used in this study.\label{tab2}}

		\begin{tabularx}{\textwidth}{
  >{\centering\arraybackslash}p{2.2cm}            
  >{\centering\arraybackslash}p{2.8cm}            
  >{\centering\arraybackslash}p{2.2cm} 
  c                % Avg Lat
  >{\centering\arraybackslash}p{2.2cm} 
}
%} % If the paper is ``preprints'', please uncomment this parenthesis.
			\toprule
			\textbf{Number} & \textbf{Date}	& \textbf{Event Peak Time (UT)}	& \textbf{GOES Class}     & \textbf{HPC Coords (Arcsec) } \\
			\midrule

1& 9 March 2011   & 23:23		& X2.14			& (183, 253)\\

2& 6 September 2011	& 22:20		    & X3.03			& (287, 118) \\

3& 25 December 2011   & 18:16		& M5.77			& (398, $-$335)\\

4& 9 March 2012   & 03:53		& M9.15			& (49, 367)\\

5& 10 March 2012	& 17:44			& X1.22			& (377, 388)\\

6& 9 May 2012   & 12:32		& M6.81			& ($-$478, 259)\\

7& 9 May 2012	& 21:05			& M5.91			& ($-$409, 245)\\

8& 23 October 2012	& 03:17			& X2.37			& ($-$798, $-$262) \\

9& 28 October 2013   & 04:41		& M7.37			& (910, 110)\\

10& 5 November 2013	& 22:12			& X4.93			& ($-$659, $-$247) \\

			\bottomrule
		\end{tabularx}
%		\isPreprints{}{% This command is only used for ``preprints''.

%} % If the paper is ``preprints'', please uncomment this parenthesis.
	% \noindent{\footnotesize{* Tables may have a footer.}}
    \label{Tablesolar}
\end{table}
\vspace{-12pt}

\begin{table}[H]
\caption{{Summary of the number of GPS station--satellite pairs, and average SZA, latitude, and longitude of the GPS stations used in the TEC calculations during each flare.}}
\label{Tablegps}
{
\begin{tabularx}{\textwidth}{
  c                % No.
  >{\centering\arraybackslash}p{3.5cm}                 % Station-Satellite Pairs
  c                % Avg SZA
  c                % Avg Lat
  c                % Avg Long
}
\toprule
\textbf{Number} & \textbf{Number of Station--Satellite Pairs}	& \textbf{Avg. SZA (°)}	& \textbf{Avg. Lat. (°)}     & \textbf{Avg. Long. (°)} \\
\midrule
1  &  1310  & 45.80 & 2.22   & 199.72 \\
2  &  4573  & 47.19 & 38.00  & 235.36 \\
3  &  1202  & 43.95 & 1.43   & 273.96 \\
4  &  624   & 46.64 & $-$17.84 & 145.18 \\
5  &  4719  & 49.88 & 36.27  & 249.25 \\
6  &  995   & 43.18 & 30.09  & 133.66 \\
7  &  6995  & 31.75 & 38.90  & 242.27 \\
8  & 666   & 43.13 & $-$22.37 & 148.58 \\
9  &  449   & 33.56 & $-$5.86  & 115.24 \\
10 & 537   & 35.70 & $-$21.24 & 182.08 \\
\bottomrule
\end{tabularx}
}
\end{table}

\section{Methods}\label{sec3}
\subsection{TEC Detrending}\label{detrsection}
\begin{figure}[H]

% First full-width image (contains 2 panels: a and b)
\includegraphics[width=0.95\linewidth]{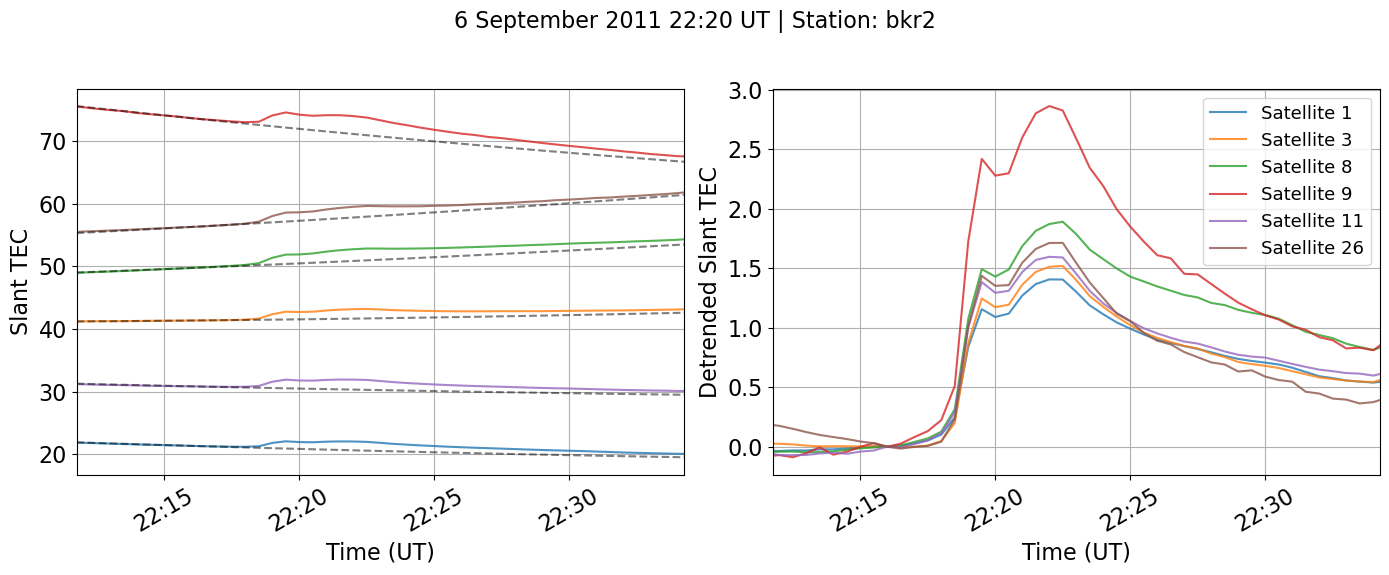}

\vspace{0.3em}
\noindent
\hspace{0.02\linewidth}
\makebox[0.45\linewidth][c]{(a)}%
\makebox[0.49\linewidth][c]{(b)}

\vspace{0.3em}

\includegraphics[width=0.95\linewidth]{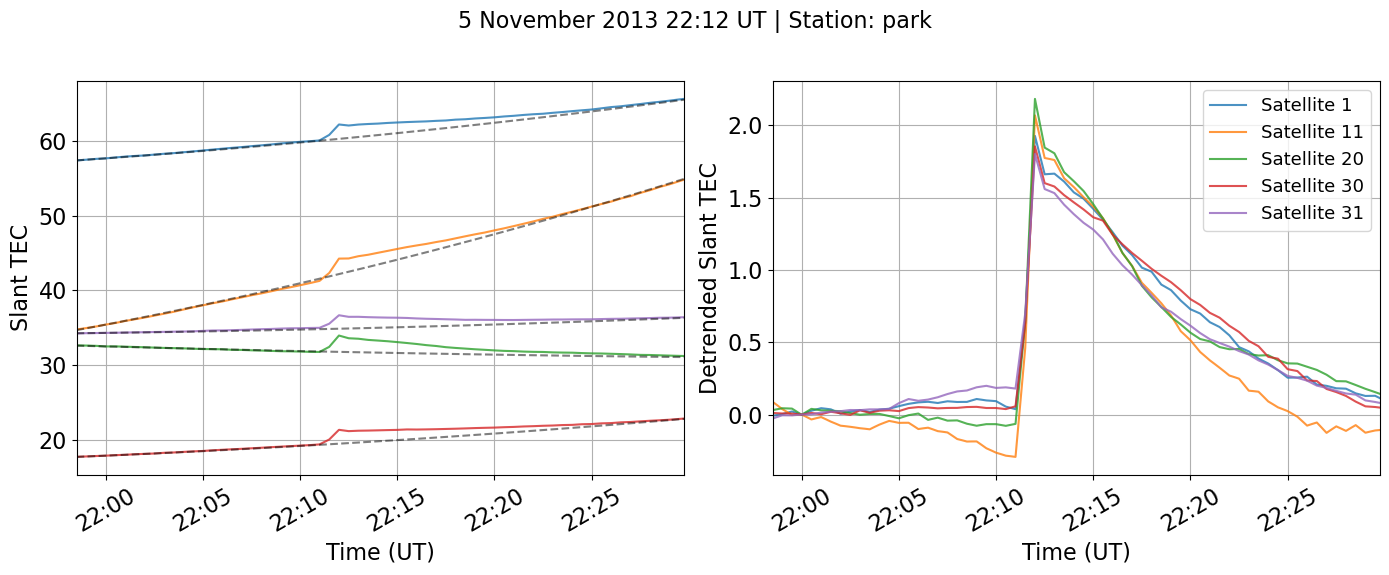}

\vspace{0.3em}
\noindent
\hspace{0.02\linewidth}
\makebox[0.45\linewidth][c]{(c)}%
\makebox[0.49\linewidth][c]{(d)}

\caption{Panel (a) shows slant TEC curves from ‘bkr2' station during Flare 3. Panel (b) shows these lines detrended after subtracting the grey dashed polynomials in panel (a). Similarly, panel (c) shows individual station--satellite slant TEC lines from ‘park' station during Flare 10 and panel (d) shows these lines after detrending. Different satellites are denoted by different colours. }
\label{detrending}
\end{figure}

As mentioned in Section \ref{ionodata}, to calculate the dynamics of the total electron content increase caused by the flares in Table \ref{Tablesolar} and make more obvious the moment of ionospheric reaction, we detrended relative TEC curves for each individual station--satellite path. {Detrending was performed using a simple polynomial fitting approach. A second-order polynomial was fitted by choosing  three reference points along each slant TEC curve outside the time range of the flare, with at least one data point taken both before and after the event, and subtracted. This technique was designed to highlight the timing of the TEC response and successfully enhanced its visibility relative to flare emission.} Examples of this process are presented in Figure \ref{detrending}. Panel (a) of Figure~\ref{detrending} shows coloured curves representing relative slant TEC measurements from multiple satellites, as observed by station ‘bkr2’ during the X3.03 solar flare on 6 September 2011 at 22:20\,UT. Grey dashed lines demonstrate the polynomials associated with satellite trajectories that were subtracted from the coloured curves to highlight a clear increase in slant TEC. Panel (b) shows the detrended slant TEC timeseries for different satellites, where we can clearly see two peaks in electron concentration caused by incident flare emission. Panels (c) and (d) show the same detrending process and results, but for station ‘park' during the X4.93 flare that occurred on 5 November 2013 (22:12\,UT). This method was applied to all of the data from the stations shown in Figure \ref{maps}. For each event, the detrended curves were averaged using a root mean squared error (RMSE) method to limit the impact of any particularly noisy or irregular station--satellite measurements. This resulted in one mean detrended TEC curve for each flare (see Section \ref{delaycalcsect}). Since this study focused on the timing of the ionospheric response rather than the magnitude of TEC enhancement, vertical TEC conversion was unnecessary, and slant TEC data sufficed.

\subsection{Calculation of Delay and Solar Emission Factors}\label{delaycalcsect}
\begin{figure}[H]
  
    \includegraphics[width=0.70\linewidth]{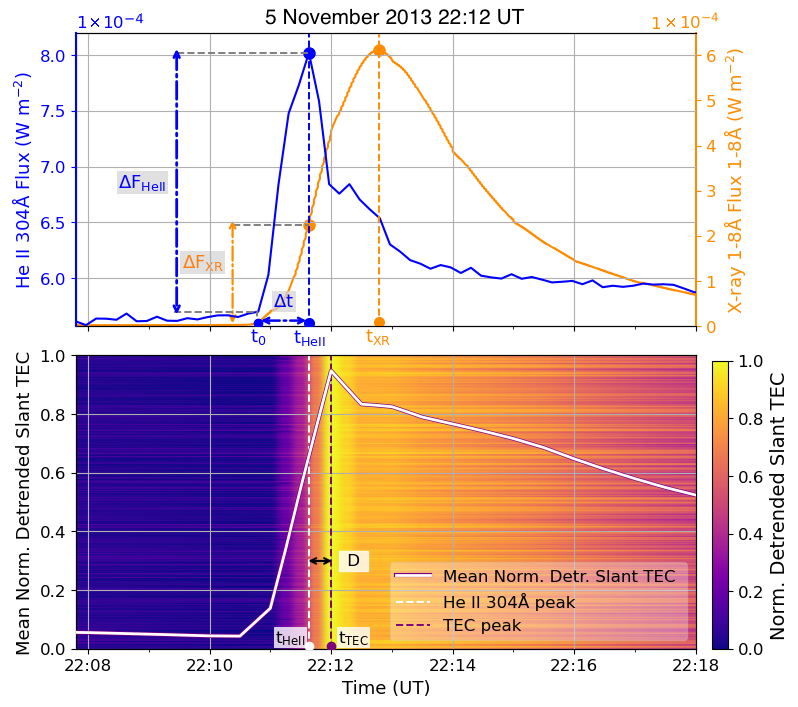}
    \caption{The top panel shows He~{\sc{II}} 304\,\AA{} (blue) and X-ray 1–8\,\AA{} (orange) lightcurves for the flare on 5 November 2013 (Flare 10). The peak times of each are shown with vertical blue (t$_{\textup{He~{\sc{II}}}}$) and orange (t$_{\textup{XR}}$) dashed lines, respectively. The flare start time is denoted by t$_{0}$, and the He~{\sc{II}} rise time is $\Delta\mathrm{\textup{t}}$. The change in He~{\sc{II}} flux ($\Delta$F$_{\textup{He~{\sc{II}}}}$) and the change in X-ray flux  ($\Delta$F$_{\textup{XR}}$) during $\Delta\mathrm{\textup{t}}$ are represented by the vertical blue and orange dash-dot arrowed lines respectively. The bottom panel shows mean normalised detrended TEC curve in white, plotted on the normalised detrended slant TECs represented by the colour bar, where each horizontal line is a station average. Delay (D) between He~{\sc{II}} and TEC response is shown between t$_{\textup{He~{\sc{II}}}}$ and t$_{\textup{TEC}}$ vertical lines. } 
    \label{delaycalc}
\end{figure}

Figure \ref{delaycalc} illustrates the delay calculation technique, using the X4.93 flare on 5 November 2013 as an example. The He~{\sc{II}} 304\,\AA{} (blue) and X-ray 1–8\,\AA{} (orange) lightcurves can be seen in the top panel of Figure \ref{delaycalc}. Their respective peak times are denoted by t$_{\textup{He~{\sc{II}}}}$ and t$_{\textup{XR}}$. As mentioned in Section \ref{solobs}, we required that these peak times be separated by no less than 1\,minute so that we could clearly distinguish the impact of He~{\sc{II}} on TEC. The increases in He~{\sc{II}} 304\,\AA{} flux ($\Delta$F$_{\textup{He~{\sc II}}}$) and X-ray flux ($\Delta$F$_{\textup{XR}}$) were calculated as the flux change from the flare start time (t$_0$) to t$_{\textup{He~{\sc{II}}}}$.

The bottom panel of Figure \ref{delaycalc} shows the normalised mean detrended TEC curve in white plotted on the corresponding normalised detrended TEC values from each station, which are represented by the colour bar. The peak time of the TEC is denoted by t$_{\textup{TEC}}$. The delay between EUV (He~{\sc{II}}) and TEC response was calculated by $\mathrm{\textup{D} = \textup{t}_{\textup{TEC}} - \textup{t}_{\textup{He~{\sc{II}}}}}$ for all flares in Table \ref{Tablesolar}. As He~{\sc{II}} emission ionises the ionospheric F-region, the D value indicates the response time of the upper ionosphere to the solar flare. The vertical brightening (yellow) illustrates how all stations react to the flare in unison, and that the mean TEC curve is a good representation of this. The observed synchronicity suggests that the influence of the solar zenith angle on the ionospheric response delay is minimal, given the precision of the measurements employed. 

As mentioned in Section \ref{intro}, the delay between the He~{\sc{II}} flare emission and the TEC response was to be assessed in relation the characteristics of the solar emissions. These included the increase in He~{\sc{II}} flux ($\Delta$ F$_{\textup{He~{\sc II}}}$), rise rate of the incident He~{\sc{II}} ($\Delta \textup{F}\mathrm{_{\textup{He~{\sc{II}}}}/}\Delta$t, where $\Delta\mathrm{\textup{t} = \textup{t}_{\textup{He~{\sc{II}}}} - \textup{t}_{\textup{0}}}$), and the ratio of X-ray flux at the He~{\sc{II}} peak ($\Delta \textup{F}\mathrm{_{\textup{XR}}/}\Delta\textup{F}\mathrm{_{\textup{He~{\sc{II}}}}}$). 

\section{Results}\label{sec4}
\subsection{{Delay Calculation}}
\begin{figure}[H]
   
    \includegraphics[width=0.76\linewidth]{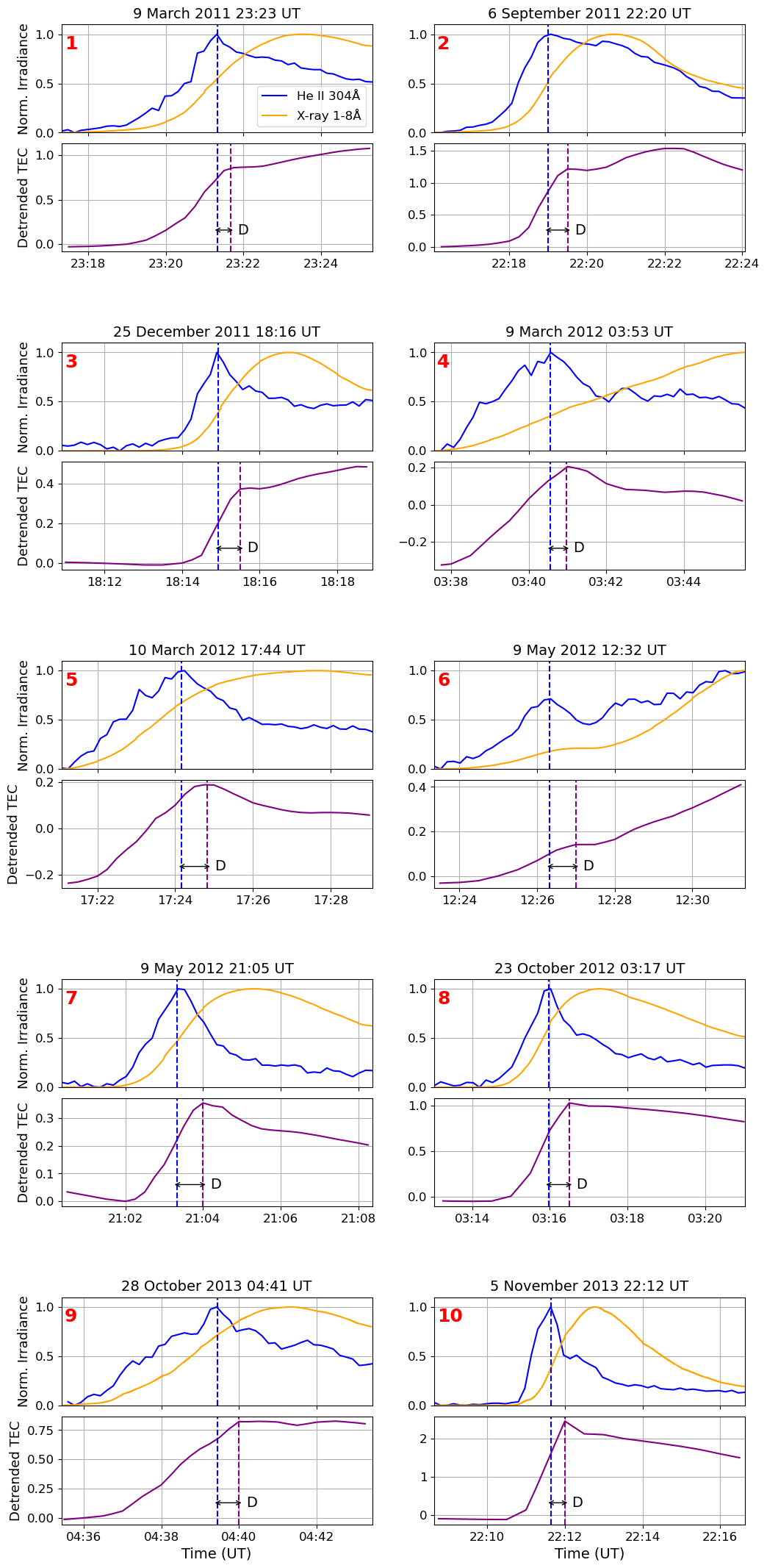}
    \caption{The top panel of each subfigure shows the normalised He~{\sc{II}} 304\,\AA{} (blue) and X-ray 1–8\,\AA{} (orange) lightcurves. The bottom panel of each subfigure shows the mean detrended TEC (purple) curves. The peak He~{\sc{II}} time is shown with blue vertical dashed line, and the peak TEC time is shown with purple vertical dashed line. The delay between solar EUV and TEC response is shown with horizontal black double arrowed line and labelled 'D'. The date and event peak time of each flare is in the title of each subfigure, and the flare numbers are shown in red. }
    \label{lcs}

\end{figure}

Figure \ref{lcs} displays the normalised flare emissions and mean detrended TEC for the 10 flares analysed in this study. The top panel of each subfigure shows the associated He~{\sc{II}} 304\,\AA{} emission in blue and X-ray 1–8\,\AA{} emission in orange, while the bottom panel presents the mean detrended TEC in purple. Figure \ref{lcs} exhibits distinct peaks in He~{\sc{II}} emission for each flare, with corresponding dynamics in the TEC response. For example, in Flares 1, 2, and 3 (9 March 2011, 6 September 2011, and 25 December 2011), TEC initially peaks in response to the He~{\sc{II}} emission, followed by a plateau or slight dip (since recombination temporarily prevails over ionisation) before rising again as the X-ray emission increases. 
While the peak in X-ray flux produces a noticeable enhancement in TEC, its impact is less significant than that of the initial He~{\sc{II}}-driven ionisation, as it affects only a narrow, low-altitude region. In some flares, we observed a double-peaked He~{\sc{II}} lightcurve, characterised by an initial peak, a subsequent decrease, followed by a rise. This ‘pre-flare' peak can also be used to measure delay, as demonstrated in Flare 6 (9 May 2012 12:32\,UT). Here, the He~{\sc{II}} emission shows an initial peak before declining and and rising again, while the TEC response mirrors this behaviour. 
The time delays (D) between the He~{\sc{II}} and corresponding TEC response for each of the flares are labelled in Figure \ref{lcs}. All measured delays were less than one minute, though with notable flare-to-flare variability. Therefore, to explore this difference, we investigated various potential contributing factors. We note, however, that the 15\,s cadence of TEC measurements limits the precision of these delay estimates. {Combined with the uncertainty in the He~{\sc{II}} observations, we estimate an uncertainty of $\sim$9\,s in our delay values.} Despite this constraint, our findings remain valuable for assessing relationships between different parameters and observed TEC delays.

\subsection{{Correlation with Heliogeophysical Parameters}}
\begin{figure}[H]
  
    \includegraphics[width=0.8\linewidth]{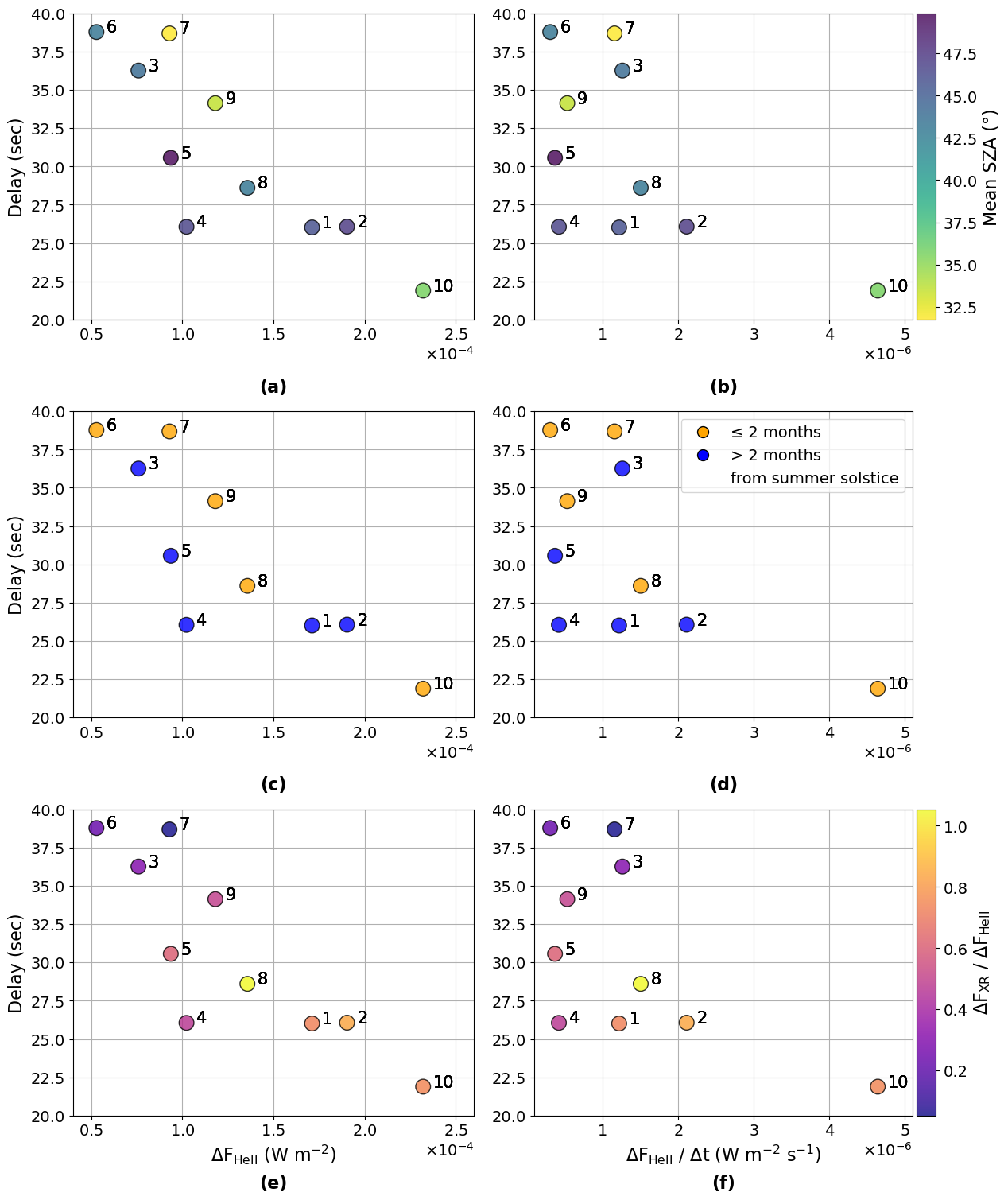}
    \caption{The left panels (a, c, e) show delay as a function of He~{\sc{II}} flux change ($\Delta$F$_{\textup{He~{\sc{II}}}}$). The right panels (b, d, f) display delay as a function of $\Delta \textup{F}\mathrm{_{\textup{He~{\sc{II}}}}/}\Delta$t. In panels (a, b), the colour scale indicates the average solar zenith angle of the GPS stations for each flare. In panels (c, d), blue represents flares that occurred more than 2\,months away from summer solstice, and orange denotes flares that occurred within 2\,months of summer solstice.
    In panels (e, f), the colour bar shows the ratio between the X-ray flux ($\Delta$F$_{\textup{XR}}$) at the peak time of He~\textup{{\sc{II}}} and the peak He~{\sc{II}} flux ($\Delta$F$_{\textup{He~{\sc{II}}}}$). The numbers 1--10 correspond to the flare numbers listed in Table \ref{Tablesolar}.}
\label{delay_factors_4grid}
\end{figure}

As we expect He~{\sc{II}} flux to be the dominant driver of TEC variations, we first analysed the relationship between the calculated delays and the characteristics of the He~{\sc{II}} 304\,\AA{} flux during the flare. The top panels of Figure \ref{delay_factors_4grid} show the relationship between delay and both the absolute increase in He~{\sc{II}} flux ($\Delta$F$_{\textup{He~{\sc{II}}}}$; (a)) and the gradient of this increase ($\Delta \textup{F}\mathrm{_{\textup{He~{\sc{II}}}}/}\Delta$t; (b)). The colour bar indicates the mean SZA values for the stations used in \textls[-15]{TEC measurements for each flare. As evident, there is a strong negative correlation} {(see e.g., \cite{Schober2018, Papageorgiou2022}, etc.)} between $\Delta$F$_{\textup{He~{\sc{II}}}}$ and delay. This is confirmed by the calculated Pearson (c$_p=-0.84$) and Spearman (c$_s=-0.87$) correlation coefficients (see Table \ref{tab:correlation_coefficients}). However, $\Delta\textup{F}\mathrm{_{\textup{He~{\sc{II}}}}/}\Delta$t exhibits a weaker negative correlation with delay, yet still moderate, with correlation coefficients of c$_p=-0.59$ and c$_s=-0.54$. {The \emph{p}-values (\mbox{Table \ref{tab:correlation_coefficients}}) for this relationship are higher than 0.05, indicating that our result is not statistically significant. This is expected as our sample of flares does not continuously cover a range of observed $\Delta\textup{F}\mathrm{_{\textup{He~{\sc{II}}}}/}\Delta$t. For a larger set of events with a broader range of He~{\sc{II}} flux rates, this correlation may be stronger and more statistically significant.} Nonetheless, these results suggest that both faster and greater increases in He~{\sc{II}} will cause ionisation of the F-region to reach a maximum in less time. This agrees with what has been previously demonstrated for X-rays and the D-region reaction \cite{basak2013effective}. The difference in correlation coefficients between delay and these two factors tells us that the overall change in He~{\sc{II}} flux during the flare has a bigger impact on the response time of the ionospheric electron content than the ‘sharpness' or impulsiveness of the incident He~{\sc{II}} emission. It can be noticed from panel (a) that in some cases, for flares with similar $\Delta$F$_{\textup{He~{\sc{II}}}}$, those with smaller SZA tend to have longer delays. However, the same trend is not exhibited in panel (b), and the correlation between delay and SZA is too weak (c$_p= 0.31$ and c$_s=0.30$) to state the influence of SZA. {Additionally, the high p-values for this relationship indicate this result is not statistically significant and therefore cannot be used to draw any conclusions. Potentially for the same reasons, we were unable to find a relationship between delay and average latitude of the stations used.}

Another geophysical factor impacting delay is the time of year the flare occurs. Panels (c) and (d) of Figure \ref{delay_factors_4grid} again display delay as a function of $\Delta$F$_{\textup{He~{\sc{II}}}}$ and $\Delta \textup{F}\mathrm{_{\textup{He~{\sc{II}}}}/}\Delta$t, respectively, with colour now highlighting seasonal variation. Flares within 2\,months of the summer solstice are represented by orange points, others are blue. For flares with comparable $\Delta$F$_{\textup{He~{\sc{II}}}}$, we observe longer delays for those events occurring nearer to the summer solstice. This pattern can be attributed to elevated temperatures during summer months, which reduce recombination rates. Because recombination rates directly influence how quickly the ionosphere reacts to a flare, slower recombination prolongs the ionisation phase in response to solar radiation. As a result, the peak ionospheric response occurs later.  

\begin{table}[H]
\centering
\renewcommand{\arraystretch}{1.2}
\setlength{\tabcolsep}{10.5pt}
\caption{Pearson and Spearman correlation coefficients between irradiance and geophysical parameters and TEC delay. {Corresponding p-values are included to indicate statistical significance.}}
\begin{tabular}{c c c c c c}
\toprule
& \multicolumn{5}{c}{\textbf{Delay vs}} \\
\cmidrule{2-6}
& \boldmath{$\Delta$}\textbf{F}\textbf{$_{\text{He\,II}}$} 
& \boldmath{$\Delta$}\textbf{F}\textbf{$_{\text{He\,II}}/\Delta \textup{t}$} 
& \textbf{SZA}
& \boldmath{$\Delta$}\textbf{F}\textbf{$_{\text{XR}}$}\textbf{/}\boldmath{$\Delta$}\textbf{F}\textbf{$_{\text{He\,II}}$}
& \textbf{GOES Class} \\
\midrule
Pearson $c_p$ & $-$0.84 & $-$0.59 & $-$0.31 & $-$0.78 & $-$0.77 \\

{\emph{p}-value} & {0.0025} & {0.071} & {0.39} & {0.0077} & {0.0087} \\
Spearman $c_s$ & $-$0.87 & $-$0.54 & $-$0.30 & $-$0.71 & $-$0.82 \\
{\emph{p}-value} & {0.0012} & {0.11} & {0.40} & {0.022} & {0.0038} \\
\bottomrule

\end{tabular}
\label{tab:correlation_coefficients}
\end{table}

While He~{\sc{II}} 304\,\AA{} emission is a dominant driver of ionisation in the ionosphere during flares, other emission lines and continua can also play a significant role in electron concentration increase. Notably, the ionisation of the D-region by X-rays can contribute to changes in TEC by up to 23\% \cite{Bekker2024}. Therefore, we additionally analysed the relationship between the measured time delays and ratio of the X-ray to He~{\sc{II}} flux increases ($\Delta$F$_{\textup{XR}}$/$\Delta$F$_{\textup{He~{\sc{II}}}}$). While the bottom panels of Figure \ref{delay_factors_4grid} show the same dependence of delay on $\Delta$F$_{\textup{He~{\sc{II}}}}$ and $\Delta \textup{F}\mathrm{_{\textup{He~{\sc{II}}}}/}\Delta$t,  the colour in panels (e) and (f) represents $\Delta$F$_{\textup{XR}}$/$\Delta$F$_{\textup{He~{\sc{II}}}}$ at the moment of the He~{\sc{II}} peak (t$_{\textup{He~{\sc{II}}}}$). A statistically significant negative correlation exists between the measured delays and $\Delta$F$_{\textup{XR}}$/$\Delta$F$_{\textup{He~{\sc{II}}}}$, which is confirmed by the correlation coefficients;  c$_p=-0.78$ and c$_s=-0.71$.
This indicates that TEC increases more rapidly when X-rays contribute more substantially to the total ionising flux.

\begin{figure}[H]
 
    \includegraphics[width=0.79\linewidth]{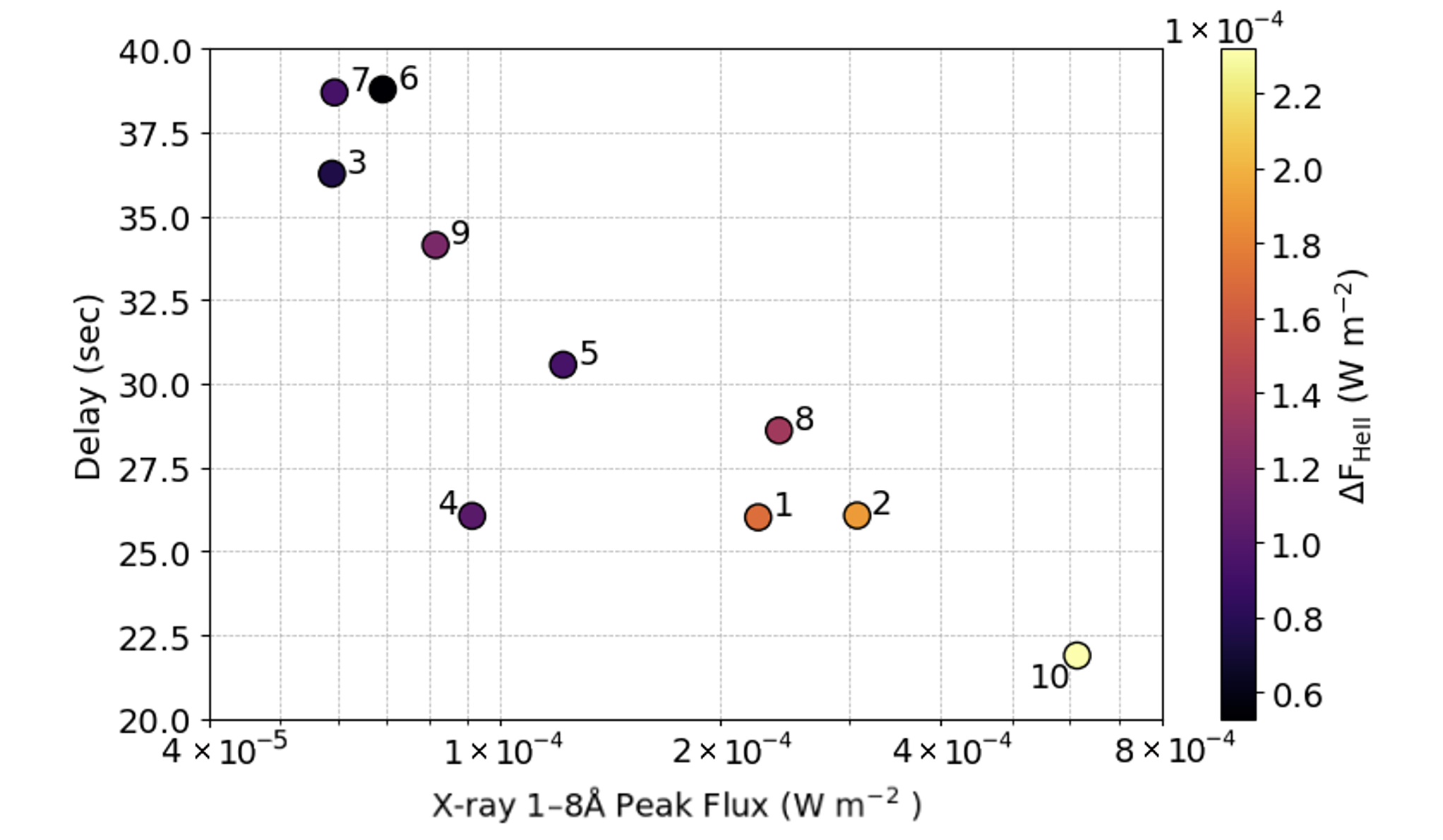}
    \caption{Delay as a function of the peak X-ray 1–8\,\AA{} flux. The colour represents the change in He~{\sc{II}} 304\,\AA{} flux for each flare. The numbers 1--10 correspond to the flare numbers listed in Table \ref{Tablesolar}.} 
    \label{delay_XRclass_He}
\end{figure}

Flares of greater magnitude typically have a greater associated increase in EUV flux (e.g., \cite{Le2007, Sreeraj2025}). For this reason, we expected to observe a relationship between the delay and the flare class. The standard classification of flare magnitude is derived from observations of the peak flux in the 1–8\,\AA{} channel from the X-ray sensor (XRS; \cite{hanser1996design}) onboard GOES. \mbox{Figure \ref{delay_XRclass_He}} displays delay as a function of X-ray 1–8 \AA{} peak flux and the colour bar represents He~{\sc{II}} 304\,\AA{} flux change. The correlation coefficients (see Table \ref{tab:correlation_coefficients}) between delay and X-ray peak flux are c$_p=-0.77$ and c$_s=-0.82$, which are both deemed a strong negative correlation. This result further confirms that the correlation with $\Delta$F$_{\textup{He~{\sc{II}}}}$ was not coincidental. Since the primary driver of TEC dynamics is the He~{\sc{II}} line, it is expected to observe slightly lower correlation coefficients between D and the X-ray flux than between D and the increase in He~{\sc{II}} flux.

\section{Discussion and Conclusion}\label{sec5}
This paper presents a study of the ionospheric F-region's sluggishness by analysing the delays between EUV solar emission, { specifically He~{\sc{II}} 304\,\AA{}}, and the corresponding changes in ionospheric electron density during 10 solar flares. {We measure delays ranging from $\sim$20–40 s, which is in close agreement with the limited previous measurements for single events from \citet{bekker2024influence} and \citet{ohare2025quasi}, who reported delays of 60 s and 30 s, respectively. The cadence of the TEC measurements causes additional uncertainty in determining response times, particularly when delays are short. When combined with the cadence limitations of the He~{\sc{II}} observations, we estimate that this introduces a typical uncertainty of approximately ±9 s in our delay measurements, corresponding to a percentage range of $\sim$23--41$\%$. Higher-cadence TEC data do exist at certain stations, but their limited spatial and temporal coverage rendered them unsuitable for this multi-event study, particularly over the SDO MEGS-A operational period. }

{Our analysis reveals a strong correlation between $\Delta$He~{\sc{II}} 304~\AA{} flux and how quickly the ionosphere responds which mirrors findings by \citet{basak2013effective} for the D-region.} Additionally, we find a moderate correlation between the rate of increase in He~{\sc{II}} 304\,\AA{} emission ($\Delta \textup{F}\mathrm{_{\textup{He~{\sc{II}}}}/}\Delta $t) and how fast we see a response in TEC, suggesting that more impulsive (sharp) flares deliver ionising radiation to the ionosphere {more quickly, prompting faster ionospheric responses. Together, these results demonstrate that both the magnitude and rate of change of He~{\sc{II}} 304\,\AA{} flux play key roles in shaping the timing of the ionospheric response.}

{Previous studies, including \citet{basak2013effective}}, 
have shown that the time delay between X-ray flare emission and the D-region response increases with SZA. {In our study, no consistent relationship between delay and SZA emerges. This may be due to various contributing factors, including season, solar cycle phase, and pre-flare ionospheric conditions. This could also be due to the small sample size, but more likely reflects the cadence limitations of the available data. Future investigations using larger datasets with better temporal resolution may provide a clearer relationship between delay and SZA.} At the same time, we identified a relationship between the delay and the time of year. For flares with similar $\Delta$F$_{\textup{He~{\sc{II}}}}$, longer delays were observed for events occurring closer to the summer solstice. This can be explained by higher electron temperatures during that period, which leads to lower recombination rates. Since the recombination rate directly determines the speed of the ionospheric response to a flare, lower recombination results in a prolonged ionisation phase under solar radiation, thereby {delaying the observed peak in TEC}.

Interestingly, we see a strong anti-correlation between delay and $\Delta$F$_{\textup{XR}}$/$\Delta$F$_{\textup{He~{\sc{II}}}}$ at the peak of the He~{\sc{II}} emission. This indicates that the contribution of X-ray radiation to overall ionospheric ionisation accelerates the increase in TEC and, consequently, leads to an earlier peak in electron density. This aligns with the findings of \citet{Bekker2024} and \citet{Nishimoto2023}. A comparable strong anti-correlation is seen between delay and GOES flare class. This is intuitive as we expect a larger increase in EUV flux for a larger class of flare, which has been observationally proven by numerous studies such \mbox{as \cite{Le2011,Sreeraj2025,Le2013,Hazarika2016,Mahajan2010,Saharan2023}}. Despite the correlation between X-ray and ultraviolet radiation during solar flares, the primary driver of electron content dynamics in the ionosphere is EUV radiation. Therefore, it is natural that the delay shows a slightly weaker correlation with X-ray flux than with He~{\sc{II}} flux.

The delays presented in this study {may be applicable in estimating recombination rates in the E- and F-regions, but further model validation is needed. We} acknowledge that such estimation of the recombination rate is more difficult for higher ionospheric regions, primarily due to the thicker ionisation altitude range and the absence of empirical two-parameter models for the F-region, analogous to those that exist for the \mbox{D-region \cite{Ryakhovsky2024, Thomson1993, Ferguson1995}}. Further work should examine a larger sample of flares and incorporate the analysis of other geoeffective flare emission lines, such as those suggested in ionospheric modelling papers \cite{Nishimoto2023}. However, this study and future investigations surrounding flare emission and the impact on the ionosphere are limited by the availability and cadence of TEC measurements and EUV observations, highlighting the need for further improvements to our infrastructure and space missions.

\vspace{6pt}

\authorcontributions{Conceptualisation, A.N.O. and S.B.; methodology, A.N.O., S.B. and H.J.G.; software, A.N.O., and H.J.G.; validation, A.N.O., S.B. and H.J.G.; formal analysis, A.N.O. and S.B.; investigation, A.N.O. and S.B.; resources, A.N.O.; data curation, A.N.O. and H.J.G.; writing---original draft preparation, A.N.O.; writing---review and editing, A.N.O., S.B., H.J.G., R.O.M.; visualisation, A.N.O. and S.B.; supervision, S.B. and R.O.M.; project administration, A.N.O. and S.B.; funding acquisition, R.O.M. All authors have read and agreed to the published version of the manuscript.}

\funding{This research was funded by the European Office of Aerospace Research and Development grant number FA8655-22-1-7044P00001. H.J.G. and R.O.M. acknowledge support from the Science and Technology Facilities Council (STFC) grant ST/X000923/1.}

\dataavailability{The SDO/EVE data are publicly available at \url{https://lasp.colorado.edu/eve/data_access/index.html} (accessed on 1 March 2025). The GOES data can be accessed from \url{https://www.ngdc.noaa.gov/stp/satellite/goes-r.html} (accessed on 1 March 2025). Data from the SOPAC network are available at \url{http://sopac-old.ucsd.edu/} (accessed on 10 March 2025).}

\acknowledgments{This research was supported by the International Space Science Institute (ISSI) in Bern, through ISSI International Team project \#24-618.}

\conflictsofinterest{The authors declare no conflicts of interest. The funders had no role in the design of the study; in the collection, analyses, or interpretation of data; in the writing of the manuscript; or in the decision to publish the results.} 
\pagebreak

\abbreviations{Abbreviations}{
The following abbreviations are used in this manuscript:
\\

\noindent 
\begin{tabular}{@{}ll}
SDO & Solar Dynamics Observatory\\
EVE & Extreme Ultraviolet Variability Experiment\\
TEC & Total Electron Content\\
EUV & Extreme Ultraviolet\\
GOES & Geostationary Operational Environmental Satellite\\
XRS & X-ray Sensor\\
GPS & Global Positioning System\\
SOPAC & Scripps Orbit and Permanent Array Center\\
GNSS & Global Navigation Satellite System\\
RMSE & Root Mean Square Error\\
HPC & Helioprojective Cartesian Coordinates\\
HEK & Heliophysics Event Knowledgebase\\

\end{tabular}
}

\begin{adjustwidth}{-\extralength}{0cm}

\reftitle{References}

% % ACS format
\isAPAandChicago{}{

}{}

%%%%%%%%%%%%%%%%%%%%%%%%%%%%%%%%%%%%%%%%%%
%% for journal Sci
%\reviewreports{\\
%Reviewer 1 comments and authors’ response\\
%Reviewer 2 comments and authors’ response\\
%Reviewer 3 comments and authors’ response
%}
%%%%%%%%%%%%%%%%%%%%%%%%%%%%%%%%%%%%%%%%%%
\PublishersNote{}
% \isPreprints{}{% This command is only used for ``preprints''.
\end{adjustwidth}
% } % If the paper is ``preprints'', please uncomment this parenthesis.
\end{document}